\documentclass[aps,prl,twocolumn,superscriptaddress]{revtex4-1}
\usepackage{amsmath}
\usepackage{graphicx}
\usepackage{bm}
\newcommand{\ket}[1]{|#1\rangle}             
\begin{document}
\title{Spin-to-orbital angular momentum conversion and spin-polarization filtering in electron beams}
\author{Ebrahim Karimi}
%
%
\affiliation{Dipartimento di Scienze Fisiche, Universit\`{a} di
Napoli ``Federico II'', Complesso Universitario di Monte S. Angelo,
80126 Napoli, Italy}
\author{Lorenzo Marrucci}
\email{marrucci@na.infn.it}
\affiliation{Dipartimento di Scienze Fisiche, Universit\`{a} di
Napoli ``Federico II'', Complesso Universitario di Monte S. Angelo, 80126
Napoli, Italy}
\affiliation{CNR-SPIN, Complesso Universitario di Monte S. Angelo, 80126
Napoli, Italy}
\author{Vincenzo Grillo}
\affiliation{CNR-Istituto Nanoscienze, Centro S3, Via G Campi 213/a,
I-41125 Modena, Italy} 
\affiliation{CNR-IMEM, Parco delle Scienze 37a, I-43100 Parma,
Italy}
\author{Enrico Santamato}
\affiliation{Dipartimento di Scienze Fisiche, Universit\`{a} di
Napoli ``Federico II'', Complesso Universitario di Monte S. Angelo, 80126
Napoli, Italy}
\begin{abstract}
We propose the design of a space-variant Wien filter for electron
beams that induces a spin half-turn and converts the corresponding
spin angular momentum variation into orbital angular momentum (OAM)
of the beam itself, by exploiting a geometrical phase arising in the
spin manipulation. When applied to a spatially-coherent input
spin-polarized electron beam, such device can generate an electron
vortex beam, carrying OAM. When applied to an unpolarized input
beam, the proposed device in combination with a suitable diffraction
element can act as a very effective spin-polarization filter. The
same approach can be also applied to neutron or atom beams.
\end{abstract}
\maketitle
Phase vortices in electronic quantum states have been widely investigated
in condensed matter, for example in connection with
superconductivity, the Hall effect, etc. Only very recently,
however, free-space electron beams exhibiting controlled phase
vortices have been experimentally generated in transmission electron
microscope (TEM) systems, using either a spiral phase plate obtained
from a stack of graphite thin films \cite{uchida10}, or a
``pitchfork'' hologram manufactured by ion beam lithography
\cite{verbeeck10,mcmorran11}. In a cylindrical coordinate system
$r,\phi,z$ with the $z$ axis along the beam axis, a vortex electron
beam is described by a wavefunction having the general form
$\psi(r,\phi,z,t) = u(r,z,t)\exp(i\ell\phi)$, where $\ell$ is a
(nonzero) integer and $u$ vanishes at $r=0$. As in the case of
atomic orbitals, $\ell$ is the eigenvalue of the $z$-component
orbital angular momentum (OAM) operator $\hat L_z=-i\partial_\phi$
(in units of the reduced Planck constant $\hbar$), and therefore an
electron beam of this form carries $\ell\hbar$ of OAM per electron
\cite{bliokh07}.
%
%
The recent experiments on electron vortex beams were inspired by the
singular optics field, in which similar phase or holographic tools have been
used in the last twenty years (see, e.g., \cite{frankearnold08} and references therein).
In optics, a recently introduced alternative approach to the
generation of vortex beams is based on the ``conversion'' of
the angular momentum variation occurring in a spin-flip process into
orbital angular momentum of the light beam, when the latter is propagating
through a suitable spatially variant birefringent plate
\cite{marrucci06prl,marrucci11}. In this
paper, we propose that a beam of electrons traveling in free space
undergoes a similar ``spin-to-orbital angular momentum conversion''
(STOC) process in the presence of a suitable space-variant magnetic
field. The same approach may work also for neutrons, or any other
particle endowed with a spin magnetic moment (e.g., atoms, ions).
Of course, in the case of electrons, as for other
charged particles, the magnetic field, besides acting on the spin,
will also induce forces that must be compensated in order to avoid
strong beam distortions or deflections. Such compensation may be
obtained by a suitable electric field, and this leads us to
conceiving the proposed apparatus essentially as a
\emph{space-variant Wien filter}. Such apparatus can be
exploited for generating vortex electron beams when a spin-polarized
beam is used as input. Conversely, if a pure vortex beam is used in
input, by means e.g.\ of a holographic method, one can use the STOC
process for filtering a single spin-polarized component of the input
beam, as we will show further below.

Let us consider an electron beam propagating in vacuum along the
$z$-axis and crossing a region of space lying between $z=0$ and $z=L$
in which it is subject to electric
and magnetic fields $\bm{E}=-\nabla\Phi$ and
$\bm{B}=\nabla\times\bm{A}$, where $\Phi$ and
$\bm{A}$ are the scalar and vector potentials, respectively.
In the non-relativistic approximation and
neglecting all Coulomb self-interaction effects (small charge
density limit), the electron beam quantum propagation and spin
evolution are generally described by Pauli's equation
\begin{equation}\label{eq:pauli}
   i\hbar\partial_t\tilde\psi = \left[\frac{1}{2m}(-i\hbar\bm{\nabla}-e \bm{A})^2
   + e\Phi - \bm{B} \cdot \hat{\bm{\mu}}\right]\tilde\psi
\end{equation}
where $\tilde\psi$ is the spinorial two-component wave-function of
the electron beam, $e=-|e|$ and $m$ are the electron charge and
mass, $\partial_t$ is the derivative with respect to the time
variable $t$, $\hat{\bm{\mu}}=-\frac{1}{2}g\mu_B\hat{\bm{\sigma}}$
is the electron magnetic moment, with $\mu_B=\hbar |e|/2m$ the
Bohr's magneton, $g\simeq 2$ the electron $g$-factor, and
$\hat{\bm{\sigma}}=(\hat\sigma_x,\hat\sigma_y,\hat\sigma_z)$ the
Pauli matrix vector.

As a first step, we consider the simpler case in which the electric
and magnetic fields are taken to be uniform, lying in the transverse
plane $xy$, and arranged as in standard Wien filters
\cite{tioukine06,grames11}, i.e. perpendicular to each other and
balanced so as to cancel the average Lorentz force, i.e.
$E_0=B_0p_c/m$ where $E_0$ and $B_0$ are the electric and magnetic
field moduli, and $p_c$ the average beam momentum. The magnetic
field $\bm{B}$ is also taken to form an arbitrary angle $\alpha$
with the axis $x$ within the $xy$ plane. For this case, we solved
the full Pauli's equation in the paraxial slow-varying-envelope
approximation for an input beam having a gaussian profile and an
arbitrary uniform input spin state
$\ket{\psi}_{in}=a_1\ket{\!\uparrow}+a_2\ket{\!\downarrow}$,
where $\ket{\!\uparrow}$ and $\ket{\!\downarrow}$ denote a
state for which the spin is parallel or antiparallel to the $z$ axis, respectively. The
complete expression of the resulting spinorial wave-function is
given in the supplemental material (SM) \cite{suppl}, while here we
summarize the main findings. The beam propagation behavior
corresponds to the well known astigmatic lensing
in the plane perpendicular to the magnetic field. More precisely,
the beam undergoes periodic width oscillations, with a spatial
period $\Lambda_2 = \pi R_c$, where $R_c = p_c/(|e|B_0)$ is the
cyclotron radius. This lensing phenomenon is also predicted by a
classical ray theory, when properly taking into account the effect
of the input fringe fields \cite{suppl}. The output spin state is
instead given by the following general expression (Eq. 4 in SM
\cite{suppl})
\begin{eqnarray}
\ket{\psi}_{out}&=&a_1\left[\cos(\delta/2) \ket{\!\uparrow} +
\sin(\delta/2) ie^{i\alpha}\ket{\!\downarrow}\right] \nonumber\\
&&+a_2\left[\cos(\delta/2) \ket{\!\downarrow} + \sin(\delta/2)
ie^{-i\alpha}\ket{\!\uparrow}\right], \label{eq:spinor}
\end{eqnarray}
where $\delta = 4\pi L/\Lambda_1$ and $\Lambda_1 = 4\pi R_c/g \simeq
2\Lambda_2$. This spinorial evolution corresponds to the classical
Larmor precession of the spin with spatial period $\Lambda_1/2$,
$\delta$ being the total precession angle. However, in addition to
the spin precession, Eq.\ (\ref{eq:spinor}) predicts the occurrence
of wavefunction phase shifts. In particular, for a
$\ket{\!\uparrow}$ or $\ket{\!\downarrow}$ input state and a total
spin precession of exactly half a turn, i.e. $\delta=\pi$ or
$L=\Lambda_1/4$, the wavefunction acquires a phase shift given by
$\pm\alpha+\pi/2$, where $\alpha$ is the magnetic field orientation
angle mentioned above and the $\pm$ sign is fixed by the input spin
orientation ($+$ for $\ket{\!\uparrow}$ and $-$ for
$\ket{\!\downarrow}$). These phase shifts can be interpreted as a
special case of \emph{geometric Berry phases} arising from the spin
manipulation \cite{shapere}.

\begin{figure}[!htbp]
    \includegraphics[width=8.5cm]{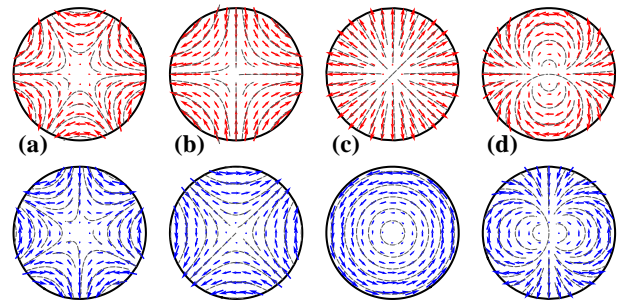}
    \caption{\label{fig:q-filter-topology} (color online) Electric (upper panels)
    and magnetic (lower panels) field $q$-filter geometries for different topological charges:
    (a) $q=-2$, (b) $q=-1$, (c) $q=1$, and (d) $q=+2$; in all cases $\beta=\pi/2$.}
\end{figure}

\begin{figure}[!htbp]
    \includegraphics[width=8.5cm]{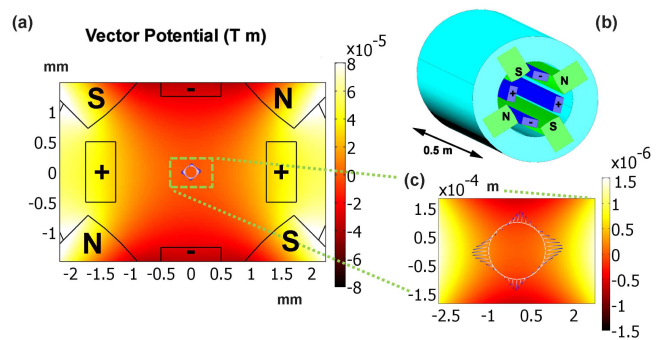}
    \caption{\label{fig:device} (color online) Electrodes and magnetic
    poles geometry of a $q$-filter with $q=-1$ (quadrupole), seen in
    cross-section (a) and in three-dimensional rendering (b).
    The filter length is set to 50 cm. In panel (a), the
    calculated vector potential $A_z$ (in false colors), which is also
    roughly proportional to the electric potential, and the projection on the $xy$
    plane of the simulated electron ray trajectories for 100 keV energy are
    also shown (see SM for details about the simulations \cite{suppl}), for a
    ring-shaped input beam with radius $r=100$ $\mu$m (ray color grows darker
    for increasing $z$). In panel (c) a zoomed-in view of the
    central region. The magnetic field at $r$ needed to obtain the tuning condition
    $\delta=\pi$ is 3.5 mT, with a corresponding electric field of
    575 kV/m. These are obtained with an electrode potential difference of $\approx 9$ kV and
    a magnetization of 135 A/mm. The fields need to be set to the design-values with
    a precision of 1 part in $10^4$.}
\end{figure}

Let us now move on to the case of a spatially variant magnetic
field. We consider multipolar transverse field geometries with
cylindrical symmetry, described by the following expression for the
magnetic field (with the vector given in cartesian components):
$\bm{B}(r,\phi,z)=B_{0}(r)(\cos{\alpha(\phi)},\sin{\alpha(\phi)},0)$,
where the angle $\alpha$ is now the following function of the
azimuthal angle:
\begin{equation}\label{eq:alpha}
  \alpha(r,\phi,z) = q\phi + \beta
\end{equation}
where $q$ is an integer and $\beta$ a constant. Clearly, such a
field pattern must have a singularity of topological charge $q$ at
$r=0$. In particular, by imposing the vanishing of the field
divergence, we find that the radial factor $B_0(r) \sim r^{-q}$,
i.e., the field vanishes on the axis for $q<0$, while it diverges
for $q>0$. In the latter case, there must be a field source on the
axis. We call ``$q$-filter'' a balanced Wien filter whose magnetic
field distribution in the beam transverse plane obeys Eq.\
(\ref{eq:alpha}). The electric field will be taken to have an
identical pattern, except for a local $\pi/2$ rotation, so as to
balance the Lorentz force. Some examples of such $q$-filter field
distributions are shown in Fig.\ \ref{fig:q-filter-topology}.
We are particularly interested in the negative $q$ geometries, which
do not require to have a field source at $r=0$. For example, the
$q=-1$ case corresponds to the standard quadrupole geometry of
electron optics, while $q=-2$ corresponds to the hexapole one. Wien
filters with such geometries have been already developed in the past
for the purpose of correcting chromatic aberrations
\cite{rose70,zach95}. Moreover, inhomogeneous Wien filters including
several multipolar terms have been also considered for the purpose
of spin manipulation, with the added advantage of obtaining a
stigmatic lensing behavior \cite{Scheinfein89}. A possible design of
the $q=-1$ filter with quadrupolar geometry is shown in Fig.\
\ref{fig:device}.
In such non-uniform field geometry we cannot solve analytically the
full Pauli's equation. However, the beam propagation is already well
described by classical dynamics and can be derived either
analytically, using a power-expansion in $r$ \cite{Scheinfein89}, or
by numerical ray tracing. In the former case, we find that to first
order the $q$-filter for $q\neq0$ is already stigmatic, i.e., it
preserves the beam circular symmetry. Only second-order corrections
introduce aberration effects \cite{Scheinfein89}. This behavior is
confirmed by our numerical ray-tracing simulations (see SM for
details \cite{suppl}), which show relatively weak higher-order
aberrations (see Fig.\ \ref{fig:device}). It should be noted that
these simulations have been performed for realistic values of the
electric and magnetic fields, as required to obtain a spin
precession of half a turn across a propagation distance of 50 cm at
a beam radius of 100 $\mu$m. These calculations are expected to
reproduce very well the electron density behavior (and spin
precession) as would be obtained from Pauli's equation. However,
Pauli's equation predicts an additional purely quantum phenomenon,
namely the geometric phase already discussed above. In a
semi-classical approximation, the geometric phase will still be
given by $\pm\alpha+\pi/2$, where $\alpha$ is however now
position-dependent and given by Eq.\ (\ref{eq:alpha}). More
specifically, neglecting the aberrations, each possible electron
trajectory within a beam is straight and parallel to the $z$ axis.
Therefore, the electrons travelling in a given trajectory will
experience a constant magnetic field of modulus $B_0(r)$ and
orientation $\alpha(\phi)$. The set of all electrons travelling at a
given radius $r$ will then undergo a uniform spin precession by
angle $\delta(r)$ and, if $\delta(r)=\pi$ (i.e., for a spin
half-turn rotation), they will also acquire a \emph{space-variant
geometric phase} given by $\pm\alpha(\phi)+\pi/2=\pm q\phi \pm\beta
+\pi/2$, with the $\pm$ sign determined by the input spin
orientation. In other words, the outgoing wavefunction acquires a
phase factor $\exp(i\ell\phi)$, with $\ell=\pm q$, corresponding to
a vortex beam with OAM $\pm q\hbar$. In a quantum mechanical
notation, spin-polarized input electrons with given initial OAM
$\ell$ passing through a $q$-filter undergo the following
transformations:
\begin{eqnarray}\label{eq:qftransf}
  \ket{\!\uparrow,\ell} & \rightarrow & \cos(\delta/2)
  \ket{\!\uparrow,\ell}+i e^{i\beta}\sin(\delta/2)
  \ket{\!\downarrow,\ell+q}\nonumber\\
  \ket{\!\downarrow,\ell}& \rightarrow & \cos(\delta/2)
  \ket{\!\downarrow,\ell}+i e^{-i\beta}\sin(\delta/2) \ket{\!\uparrow,\ell-q}
\end{eqnarray}
where the ket indices now specify both the spin state (arrows) and
the OAM eigenvalue.

Equations (\ref{eq:qftransf}) show that, in passing through the
$q$-filter, a fraction $f=\sin^2(\delta/2)$ of the electrons in the
beam will flip their spin and acquire an OAM $\pm q\hbar$, while the
remaining fraction $1-f=\cos^2(\delta/2)$ will pass through the
filter with no change. When $L=\Lambda_1/4$, then $\delta=\pi$ and
all electrons are spin-flipped and acquire the corresponding OAM. In
the specific case $q=1$, the spin angular momentum variation for the
electrons undergoing the spin inversion is exactly balanced by the
OAM variation, so that the total electron angular momentum remains
unchanged in crossing the filter. This is the pure ``spin-to-orbital
conversion'' STOC process mentioned in the introduction, and it
occurs for $q=1$ because this geometry is rotationally invariant and
therefore no angular momentum can be exchanged with the field
sources in the filter. In the $q\neq1$ case, the input spin still
controls the sign of the OAM variation, but the total beam angular
momentum is not conserved and some angular momentum is exchanged
with the field sources. We note that this OAM variation can be also
explained as the effect of the spin-related magnetic-dipole force
acting on the electrons within the magnetic field gradients, as more
fully discussed in the SM \cite{suppl}.

The ``tuning'' condition $L=\Lambda_1/4$ or $\delta=\pi$ can be
achieved in principle for a given radius $r$ by adjusting the
strength of the magnetic and electric fields or the device length
$L$. Since the precession angle $\delta$ is $r$-dependent, however,
this tuning condition can be applied to the entire beam only if it
is shaped as a ring, i.e. with all electron density peaked at a
given radius $r$. Vortex beams with OAM $\ell\neq 0$ typically have
a doughnut shape, so they approximate a ring fairly well. On the
other hand a gaussian input beam (with $\ell=0$) cannot be fully
transformed, as $\delta=0$ at $r=0$, where the beam has the maximum
density. In such cases, only a fraction $f$ of the electrons would
be converted.

\begin{figure}[!htbp]
    \includegraphics[width=8cm]{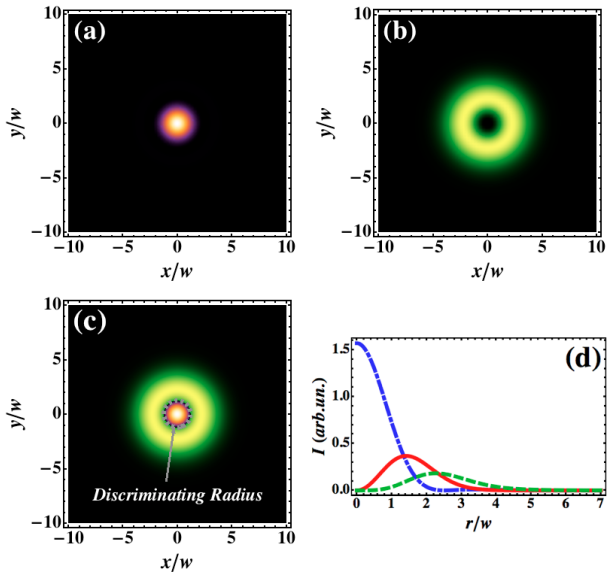}
    \caption{\label{fig:iris} (color online) Electron beam profiles in the far field of the
    $q$-filter for the $\ell=0$ component (panel a) and the $\ell=2$
    component (panel b), and a possible discriminating iris radius (panel c)
    $r=w$ to be used to separate them
    in order to make a spin-polarization filter. Panel (d) shows the intensity profiles
    of the same components ($\ell=0$ - blue dot-dashed line, $\ell=2$ - green dashed line)
    and of the possible residual $\ell=1$ component for an untuned $q$-filter (red solid
    line). $w$ is the gaussian beam waist radius in the far-field plane. A realistic
    value for the iris radius is of the order of several tens of microns (obtained
    by setting the aperture some distance after the focal plane of the second condensor)}
\end{figure}

So far we have assumed a spin-polarized input beam. However, high
brightness (i.e., spatially coherent) spin-polarized electron beams,
suitable for high-resolution TEM applications, are not so easily
available. State-of-the-art spin-polarized sources may achieve a
brightness of 10$^7$ A cm$^{-2}$ sr$^{-1}$ and a polarization purity
of up to 90\% \cite{yamamoto11} (and the source decays with time due
to laser-induced damage). It is interesting then to analyze the
effect of the $q$-filter on a initially unpolarized electron beam,
having arbitrary initial OAM $\ell$. Such input can be simply viewed
as a statistical mixture in which 50\% of the electrons are in the
state $\ket{\!\uparrow,\ell}$ and 50\% in the state
$\ket{\!\downarrow,\ell}$. After passing through a tuned $q$-filter,
the beam becomes a 50-50 mixture of states
$\ket{\!\downarrow,\ell+q}$ and $\ket{\!\uparrow,\ell-q}$, for which
spin and OAM are correlated (if the $q$-filter is not tuned, the
fraction of converted electrons decreases to $f/2$ in each spinorbit
state, and there will be a residual $1-f$ fraction of electrons in
states $\ket{\!\uparrow,\ell}$ and $\ket{\!\downarrow,\ell}$). As we
discuss now, this spin-OAM correlation can be exploited for making
an effective electron beam spin-polarization filter. Such filter
requires four basic elements in sequence (as shown in Fig.\ 2 of SM
\cite{suppl}): (i) an OAM manipulation device, such as a fork
hologram \cite{verbeeck10,mcmorran11}, to set $\ell\neq0$; (ii) a
$q$-filter with $q=\ell$, generating a mixture of electrons in
states $\ket{\!\downarrow,2\ell}$ and $\ket{\!\uparrow,0}$; (iii) a
free propagation (or imaging) stage that allows these two states to
develop different radial profiles by diffraction, because of their
different OAM values; (iv) a circular aperture for finally
separating the two states. In particular, in stage (iii) state
$\ket{\!\downarrow,2\ell}$ will acquire a radial doughnut
distribution, as in Laguerre-Gaussian modes with OAM $2\ell$, which
vanishes close to the beam axis as $r^{2\ell}$, while state
$\ket{\!\uparrow,0}$ will become approximately gaussian, with
maximum intensity at the beam axis, as shown in Figs.\
\ref{fig:iris}a,b. Therefore, a suitable iris (Fig.\
\ref{fig:iris}c) will select preferentially the electrons in the
fully polarized state $\ket{\!\uparrow,0}$. An optical OAM sorter
exploiting a similar approach has been demonstrated recently
\cite{karimi09apl}.
A specific calculation for the case $|q|=|\ell|=1$ and an iris
radius equal to the beam waist $w$ in the ``far field'' yields a
transmission efficiency of our device of 55.5\% (not including the
losses arising in the OAM-manipulation device) and a polarization
degree $(I_{\uparrow}-I_{\downarrow}) /
(I_{\uparrow}+I_{\downarrow})$, where $I_{\uparrow,\downarrow}$ are
the two spin-polarized currents, of $\sim97.5\%$. Higher degrees of
polarization can be obtained at the expense of the efficiency by
reducing the iris diameter or by employing higher $q$ values (or
vice versa). It is worth noting that this apparatus works also with
a partially tuned $q$-filter, as in this case the unmodified
electron beam component is left in the initial OAM state $\ell=q$
and therefore is also cut-away by the iris. An untuned $q$-filter
will however have an efficiency reduced by the factor
$f=\sin^2(\delta/2)$. The aberrations introduced by the $q$-filter,
even if left uncorrected, might also affect its efficiency but not
its main working principle, as this is based on the vortex effect,
which is protected by topological stability. Finally, the possible
spin depolarization effect of fringe fields can be neglected if the
length-to-gap ratio of the filter is large enough
\cite{suppl,Scheinfein89}.

We note that the spin-filter application discussed above is a new
counterexample of the old statement by Bohr that free electrons
cannot be spin-polarized by exploiting magnetic fields, due to
quantum uncertainty effects \cite{darrigol84,batelaan97,gallup01}.
The reason why we can overcome Bohr's arguments is essentially that
we do not use the magnetic forces directly to obtain the separation,
but take advantage of quantum diffraction itself, as also proposed
recently in Ref.\ \cite{mcgregor11} (see SM for a fuller discussion
\cite{suppl}).

In conclusion, we believe that the $q$-filter device described in
this paper can be manufactured relatively simply, for applications
in standard electron beam sources such as those used in TEMs or
other kinds of electron microscopes. In combination with current
field-effect unpolarized electron sources, such filter might provide
a spin-polarized source with a brightness $\sim10^9$ A cm$^{-2}$
sr$^{-1}$, about two orders of magnitude higher than the current
state of the art. This result, if it will be proved practical
enough, may open the way to a spin-sensitive atomic-scale TEM, e.g.
suitable for investigating complex magnetic order in matter or for
spintronic applications.

We acknowledge the support of the FET-Open Program within the
7$^{th}$ Framework Programme of the European Commission under Grant
No. 255914, Phorbitech.

\section*{Appendix: Supplementary Material}

\subsection{Gaussian electron beam in homogeneous Wien filter}
Let us consider an electron beam propagating in vacuum along the
$z$-axis, subject to transverse and orthogonal electric and magnetic
fields $\bm{E}$ and $\bm{B}$ lying in the $xy$ plane. In this
Section, we assume the two fields to be spatially uniform, as in
standard Wien filters, but we will release this assumption later on.
If the magnetic field direction makes an angle $\alpha$ with the
$x$-axis, we may write $\bm{E}=E_{0}(\sin{\alpha},-\cos{\alpha},0)$
and $\bm{B}=B_{0}(\cos{\alpha},\sin{\alpha},0)$. As scalar and
vector potentials, we may then take
$\Phi=-E_{0}(x\sin{\alpha}-y\cos{\alpha})$ and
$\bm{A}=B_{0}\left(0,0,y\cos{\alpha}-x\sin{\alpha}\right)$,
respectively. In the non-relativistic approximation and neglecting
all Coulomb self-interaction effects (small charge density limit),
the electron beam propagation is described by Pauli's equation
\begin{equation}\label{eq:pauli}
   i\hbar\partial_t\tilde\psi = \left[\frac{1}{2m}(-i\hbar\bm{\nabla}-e \bm{A})^2
   + e\Phi - \bm{B} \cdot \hat{\bm{\mu}}\right]\tilde\psi
\end{equation}
where $\tilde\psi$ is the spinorial two-component wave-function of
the electron beam, $e=-|e|$ and $m$ are the electron charge and
mass, $\partial_t$ is the derivative with respect to the time
variable $t$, $\hat{\bm{\mu}}=-\frac{1}{2}g\mu_B\hat{\bm{\sigma}}$
is the electron magnetic moment, with $\mu_B=\hbar |e|/2m$ the
Bohr's magneton, $g\simeq 2$ the electron $g$-factor, and
$\hat{\bm{\sigma}}=(\hat\sigma_x,\hat\sigma_y,\hat\sigma_z)$ the
Pauli matrix vector.

We seek a monochromatic paraxial-wave solution with average linear
momentum $p_c$ and average energy $E_c=p_c^2/2m$ in the form
$\tilde\psi(x,y,z,t)=\exp[i(p_c z-E_c t)/\hbar]\tilde u(x,y,z)$,
where $\tilde u(x,y,z)$ is taken to be a slow-varying envelope
spinor field. Inserting this \textit{ansatz} in Eq.~(\ref{eq:pauli})
and neglecting small terms in the $z$ derivatives of $\tilde u$, we
obtain the paraxial Pauli equation
\begin{equation}\label{eq:utilde}
   \partial_z \tilde u = \frac{i}{2k_c}\left(\nabla^2_\perp - \frac{e^2}{\hbar^2}A^2+ \frac{2m}{\hbar^2}
   \bm{B}\cdot\hat{\bm{\mu}}\right)\tilde u
\end{equation}
where $\nabla^2_\perp=\partial_x^2+\partial_y^2$ is the Laplacian in
the beam transverse plane and $k_c=p_c/\hbar$ is the De Broglie
wavevector along the beam axis. In deriving Eq.~(\ref{eq:utilde}) we
set $E_0 = p_c B_0/m$, so as to cancel out the average Lorentz force
on the beam.

Equation (\ref{eq:utilde}) determines the evolution of the spinor
field $\tilde u$ along the axis $z$. This equation can be solved
analytically for the case of a gaussian input beam, as given by
$\tilde u(r,\phi,0)=\tilde a\exp[-r^2/w_0^2]$ for $z=0$, where $w_0$
is the beam waist and $\tilde a=(a_1,a_2)$ the input spinor,
corresponding to an arbitrary spin superposition
$\ket{\psi}=a_1\ket{\uparrow}+a_2\ket{\downarrow}$. A
straightforward calculation shows that the solution is given by
\begin{equation}\label{eq:utildesol}
   \tilde u(r,\phi,z)=G(r,\phi,z)\hat M(z) \tilde a
\end{equation}
where $\hat M(z)$ is the matrix
\begin{equation}\label{eq:M}
  \hat M(z) = \begin{pmatrix}
             \cos\frac{2\pi z}{\Lambda_1} & i e^{-i\alpha}\sin\frac{2\pi z}{\Lambda_1}\\
             i e^{i\alpha}\sin\frac{2\pi z}{\Lambda_1} & \cos\frac{2\pi z}{\Lambda_1}
           \end{pmatrix}
\end{equation}
where $\Lambda_1=(2\pi\hbar^2 k_c)/(m g \mu_B B_0)$ and the scalar
gaussian factor $G(r,\phi,z)$ is given by
\begin{equation}\label{eq:G}
   G(r,\phi,z) = g_0(z)e^{i k_c r^2\left(\frac{\cos^2(\alpha-\phi)}{2q_1(z)}
   +\frac{\sin^2(\alpha-\phi)}{2q_2(z)}\right)}
\end{equation}
with
\begin{eqnarray}
  g_0(z) &=& \frac{\sqrt{\pi}\, z_R \, e^{-\frac{i}{2}\arctan{\left(\frac{z}{z_R}\right)}}
  \,\,e^{-\frac{i}{2}\arctan\left(\frac{\Lambda_2}{\pi z_R}\tan\frac{\pi z}{\Lambda_2}\right)}}{(z_R^2+z^2)^\frac{1}{4}\left(\pi^2 z_R^2\cos^2\frac{\pi z}{\Lambda_2}
  +\Lambda_2^2\sin^2\frac{\pi z}{\Lambda_2}\right)^\frac{1}{4}}, \label{eq:g}\\
   q_1(z) &=& z - i z_R , \label{eq:q1}\\
   q_2(z) &=&\frac{\Lambda_2}{\pi}\left(\frac{\Lambda_2\sin(\frac{\pi z}{\Lambda_2})-i \pi z_R\cos(\frac{\pi z}{\Lambda_2})}
              {\Lambda_2\cos(\frac{\pi z}{\Lambda_2})+i \pi z_R\sin(\frac{\pi z}{\Lambda_2})}\right), \label{eq:q2}
\end{eqnarray}
where $z_R=\frac{1}{2}k_c w_0^2$ and $\Lambda_2=\frac{\pi\hbar
k_c}{e B_0}=g\Lambda_1/4\simeq\Lambda_1/2$.

The matrix $\hat M(z)$ given in Eq.\ (\ref{eq:M}) corresponds to
Eq.\ 2 in the main paper and describes the magnetic-field-induced
Larmor spin precession during propagation and the associated
geometric phases discussed in the main paper. The precession length
$\Lambda_1$ corresponds to two full spin rotations.

In addition to the spin dynamics, from Eq.~(\ref{eq:q2}), we find
that the wavefront complex curvature radius $q_2(z)$ in the
direction perpendicular to the magnetic field $\bm{B}$ changes
periodically with spatial period $\Lambda_2$. These oscillations
correspond to the well known astigmatic lensing effect of the Wien
filter.

\subsection{Space-variant Wien filter and ray-tracing simulations}
In this Section, we consider ``$q$-filter'' field geometries with
cylindrical symmetry, described by the following expression for the
magnetic field (with the vector given in cartesian components):
\begin{equation} \label{eq:B}
\bm{B}(r,\phi,z)=B_{0}(r)(\cos{\alpha(\phi)},\sin{\alpha(\phi)},0)
\end{equation}
where the angle $\alpha$ is now the following function of the azimuthal angle:
\begin{equation}\label{eq:alpha}
  \alpha(r,\phi,z) = q\phi + \beta
\end{equation}
where $q$ is an integer and $\beta$ a constant. As discussed in the
main paper, such a field pattern must have a singularity of
topological charge $q$ at the beam axis $r=0$. In particular, by
imposing the vanishing of the field divergence, we find that the
radial factor $B_0(r) \sim r^{-q}$, i.e., the field vanishes on the
axis for negative values of $q$, while it diverges for positive
values of $q$. In the latter case, there must be a field source on
the axis. The electric field will be taken to have an identical
pattern, except for a local $\pi/2$ rotation, so as to balance
everywhere the Lorentz force. We are particularly interested in the
negative $q$ geometries, which do not require to have a field source
at the beam axis. For example, the $q=-1$ case corresponds to the
standard quadrupole geometry, while $q=-2$ corresponds to an
hexapole one.

The electron propagation problem in such non-uniform field can only
be solved approximately or numerically. A second-order geometrical
optics solution is reported in Ref.\ \cite{scheinfein89}, which
includes also a detailed analysis of spin precession but without
considering the geometric phase effect (as well as the magnetic
dipole forces). This analysis shows that, to first order, a balanced
Wien filter having quadrupole or hexapole geometry but vanishing
dipole term (as in our $q$-filter) is already stigmatic, i.e. it
preserves the cylindrical symmetry without need for further
corrections. The only beam lensing or distortion effects enter as
higher-order aberrations (i.e., the $G$ term in Eq.\ (31) of Ref.\
\cite{scheinfein89}).

To further analyze the behavior of our $q$-filter, we have performed
ray-tracing simulations of the electron propagation for a quadrupole
geometry ($q=-1$). The magnetic and electric fields have been
calculated using COMSOL finite-elements simulation (www.comsol.com).
The ray-tracing routines therein contained have been used to assess
the shape of the beam at the exit of the filter starting from an
input beam having 100 keV of enery that is shaped as a ring, with a
radius $r$ of 100 $\mu$m. The magnetic field at $r$ needed to obtain
the tuning condition $\delta=\pi$ is 3.5 mT, with a corresponding
electric field of 575 kV/m. These are obtained in our geometry with
an electrode potential difference of $\approx 9$ kV and a
magnetization of 135 A/mm. The fields need to be set to the
design-values with a precision of 1 part in $10^4$. The electron
velocity to be used as initial conditions in the simulations must be
obtained from a model of the fringe fields. For the simplest
description of the input fringe fields, i.e. the so-called sharp
cut-off fringing field (SCOFF) model, the transverse velocity
components remain constant and nil, i.e. $v_x = v_y = 0$, while the
$z$-component $v_z(x,y)$ ``just inside'' the filter is given by the
following expression:
\begin{equation} \label{eq:initialv}
v_z (x,y) = v_0 - \frac{e\Phi(x,y)}{mv_0} = v_0 -
\frac{eA_z(x,y)}{m}
\end{equation}
where $v_0$ is the velocity outside the filter. This condition is
derived from the conservation of energy and also ensures the
conservation of the canonical momentum of the electrons entering the
Wien filter and hence the conservation of their wavefront
orientation.
\begin{figure}[!htbp]
    \includegraphics[width=8cm]{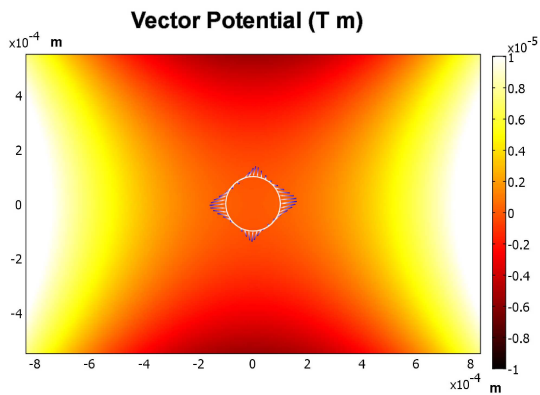}
    \caption{\label{fig:aberrations} Projection on the $xy$ plane of the
    simulated trajectories of the electrons starting from a circle of
    radius $r=100$ $\mu$m. The trajectories are shown with a color
    growing darker as $z$ is increased.}
\end{figure}
The results of ray-tracing simulations are reported in Fig.\ 2 of
the main paper and in Fig.\ \ref{fig:aberrations} of the present
manuscript, which show how a circle of electron positions
representative of the input beam becomes deformed during the
propagation into a quadrupole-lobed shape, due to second- or
higher-order aberrations. For optimizing the performances of our
proposed device, in particular in the spin-polarization filter
application, these aberrations should probably be compensated by
additional electron optics (e.g., octupole electrostatic lenses).
Nevertheless, even for no aberration compensation, the vortex effect
giving rise to the radial separation of the two spin components
discussed in the spin-polarization application is expected to be
preserved, owing to its topological stability.

\subsection{Magnetic dipole forces and angular momentum}
We now evaluate the effects of the force associated with the
magnetic dipole of the electrons in the magnetic field gradient of
the Wien filter. This is the force used in Stern-Gerlach experiments
and the only spin-dependent force acting on the electrons, so it is
important to assess its effects in detail. At any given point inside
the filter, assuming that the magnetic dipole of the electron is
$\bm{\mu}=-g\mu_B \bm{S}$, where $\bm{S}$ is the corresponding spin
vector, the associated magnetic force is
\begin{equation}
\bm{F}=\nabla(\bm{\mu}\cdot\bm{B})
\end{equation}
where the gradient must be taken while keeping $\bm{\mu}$ constant, i.e.
acting only on the magnetic field coordinates. For our classical
estimate, the magnetic dipole evolution can be described by the
precession equation
\begin{equation}
\frac{d\bm{\mu}}{dt}=\frac{g|e|}{2m}\bm{B}\times\bm{\mu}
\end{equation}
Assuming a starting magnetic moment parallel to the $z$ axis, the magnetic dipole will rotate
around the local magnetic field direction. In the approximation in which aberrations are neglected,
each electron travels along a parallel ray with constant transverse coordinates, and hence sees a
uniform magnetic field, given by Eq.\ (\ref{eq:B}). Therefore, the magnetic dipole at any given point
inside the filter will be given by the following expression:
\begin{eqnarray}
\bm{\mu}(r,\phi,z)&=& \mu_B
[-\sin\delta(r,z)\sin{\alpha(\phi)},\sin\delta(r,z)\cos{\alpha(\phi)},\nonumber\\
&& \cos\delta(r,z)] \label{eq:mu}
\end{eqnarray}
where $\delta(r,z)$ is the total spin precession angle at $z$, for a trajectory at radius $r$.
Taking the scalar product between $\bm{\mu}$
and $\bm{B}$ and then the gradient with respect to the coordinates of $\bm{B}$ only, we
find that the magnetic dipole force is only directed along the azimuthal direction and given by
\begin{equation} \label{eq:F}
\bm{F}=\hat{\bm{\phi}} q\frac{\mu_B B_0(r)}{r}\sin\delta(r,z)
\end{equation}
where $\hat{\bm{\phi}}$ denotes the unit vector along the azimuthal direction and we have used
Eq.\ (\ref{eq:alpha}) for taking the $\phi$ derivative.

Equation (\ref{eq:F}) can be used for two purposes. First, we can
verify that the ray deflection inside the Wien filter due to this
force is negligible. Indeed, integrating the force in the time $T$
needed for a full $\pi$ rotation of the spin (or spin flip),
we obtain the overall momentum change
\begin{eqnarray}
\Delta p_{\phi}&=&\int_0^{T} F_{\phi}(t)dt = \int_0^{\pi}
F_{\phi}(\delta)\left(\frac{d\delta}{dt}\right)^{-1}d\delta\nonumber\\
&& = \frac{2m}{g|e|B_0}\int_0^{\pi} F_{\phi}(\delta)d\delta
=\frac{2q\hbar}{gr}
\end{eqnarray}
We can see from this equation that the momentum variation is of the
same order as the quantum uncertainty in momentum of the electron
wavefunction, taking into account that the position uncertainty is
of the same order of the beam size $r$ (as it is also expected based
on Bohr impossibility theorem, as discussed further below). This
momentum change produces negligible effects in the electron
propagation across the filter, although it instead becomes relevant
in the far field. Indeed, it is
precisely this force that produces the orbital angular momentum
change at the basis of our spin selection. Indeed, we will now show
that the torque associated with the magnetic dipole force is that
responsible for the OAM change of the electrons. The $z$-component
torque of this force is given by
\begin{eqnarray}
M_z &=& (\bm{r}\times\bm{F})_z = q\mu_B B_0(r)\sin\delta =
q(\bm{\mu}\times\bm{B})_z \nonumber\\
&& = -q\frac{dS_z}{dt}
\end{eqnarray}
where $S_z$ is the $z$ component of the electron spin angular
momentum $\bm{S}$. Since this torque acts on the OAM of the
electrons $L_z$, we find the following angular momentum coupling
law:
\begin{equation}
\frac{dL_z}{dt} = -q\frac{dS_z}{dt}
\end{equation}
Once integrated for the entire flight time, corresponding to a full
spin flip, this coupling law returns the same result reported in the
main paper for the OAM variation induced by the spin flip.

This correspondence proves that the effect, and the only effect, of
the magnetic dipole force acting on the electrons is the change of
OAM already considered by including the geometric phase shift in the
wavefunction emerging from the filter. Therefore, the only
significant effect of this force appears in the far field and is
exactly the effect that we propose to utilize to separate electrons
according to their spin.

\subsection{Fringe field effects}
The actual fringe fields depend on the electrodes and magnetic poles geometry.
It is however safe to assume that the fringe fields extend only for a longitudinal
distance along the $z$ axis that is comparable to the transverse aperture of the
Wien filter, and therefore much smaller than the filter length, in the large
length-to-gap ratio limit.
Moreover, the transverse components of the fringe fields can be taken to have
exactly the same multipolar geometry as the fields inside the Wien, and hence
will give rise to similar effects both in lensing and spin precession. Their overall
effect is therefore only to change the effective length of the filter. However,
in addition there will also be a longitudinal $z$ component that must be taken
into account.
In the lensing effect, only the $z$ component of the electric field plays a role,
by ensuring the validity of the condition (\ref{eq:initialv}). The $z$ component
of the magnetic fields does not affect the ray propagation, since it exerts a
vanishing Lorentz force. However, the latter may affect the spin precession.
If we assume that the input electrons have a spin parallel to the $z$ axis, then
the $z$ component of the fringe magnetic field does nothing. However,
the spin may be somewhat rotated away from the $z$ axis by the transverse
fringe magnetic field, and in this case the longitudinal component will induce
some non-uniform precession around the $z$ axis that may partly depolarize
the beam. This effect will however be proportional to the time spent by the
particles in the fringe field region, and hence will be negligible in the large
length-to-gap ratio limit.

We may also consider the idealized limit of an extremely thin fringe
field region in which the electric and magnetic fields vary rapidly
from the vanishing values they have away from the Wien filter to the
values they acquire well inside the filter, which then remain
perfectly constant. This is the SCOFF limit mentioned above. In
particular, in the SCOFF limit the vector potential $\bm{A}$ must go
from zero to the value $\bm{A}=[0,0,A_z(x,y)]$ describing the
multipolar magnetic field inside the filter. This rapid transition
does not give rise to any magnetic field, because it is curl-free.
Therefore, the SCOFF limit has vanishing magnetic fringe fields and
therefore absolutely no effect on the spin, nor on the lensing
behavior of the filter. On the other hand, in the SCOFF limit the
electric potential must go from zero to the nonzero value
$\Phi(x,y)$ describing the multipolar electric field inside the
filter. This in turn implies the presence of a very strong
longitudinal electric fringe field that is responsible for the
change of velocity given by Eq. (\ref{eq:initialv}). We stress that
this idealized SCOFF description of the fringe fields needs not
being realistic, but it is sufficient to identify the ineliminable
effects that the fringe field impose to the electrons, and therefore
the possible fundamental limitations that may arise from them.

\subsection{Layout of a spin-polarized TEM microscope}
\begin{figure}[!htbp]
    \includegraphics[width=8cm]{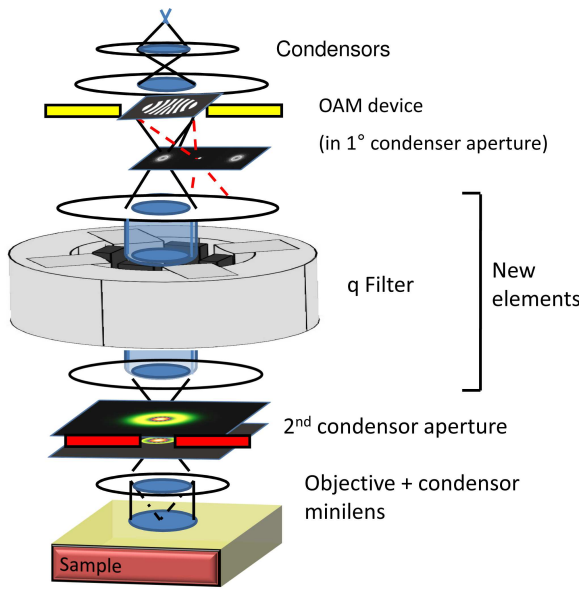}
    \caption{\label{fig:spinTEM} Sketch of a possible TEM
microscope incorporating the proposed spin-polarization filter based
on the $q$-filter.}
\end{figure}

We report in Fig.\ \ref{fig:spinTEM} the sketch of a possible TEM
microscope incorporating the proposed spin-polarization filter based
on the $q$-filter. Such a device could be used to perform
spin-sensitive TEM microscopy on magnetic materials or spintronic
devices.
It includes the four stages discussed in the main paper, i.e. (i)
the fork hologram to set the initial OAM, (ii) the $q$-filter to
couple OAM to spin, (iii) an imaging stage to go in the Fourier
plane, (iv) an aperture to select a single spinorbit component, as
well as the additional electron optics needed for imaging and
collimating the beam. The input aperture of the system is defined by
the fork hologram mask and has a typical radius of few microns
\cite{verbeeck10}. The output iris needed to select a single spin
polarization can be set to a radius of few tens of microns, by
carefully selecting the aperture plane position with respect to the
focal plane. It will be presumably convenient to work slightly off
focus to obtain larger beam waists without the need to introduce
additional magnifying optics.

We note here in passing that in the new generation of TEM
microscopes with aberration-corrected probe a larger number of
lenses are available for beam control, and in particular a pair of
hexapoles. It could be worth then exploring the possibility of using
directly these lenses as an effective $q$-filter, without
introducing additional optical elements. In this configuration, the
magnetic hexapoles with opposite sign could be used to produce the
polarization, while also compensating the introduced aberration.
However stability issues might arise for large fields and the effect
of fringing fields on the spin may also not be negligible, in this
case, as the length-to-gap ratio would not be very large. Therefore,
further analysis and simulations will be necessary in order to
assess this possibility.

\subsection{The role of electron diffraction}
In our treatment we fully included the diffraction effects only in
the free propagation (through a lensing system) taking place after
the $q$-filter device. Indeed, this stage is where the correlation
between OAM and radial profile develops and is then exploited for
separating the different spin components. This free propagation,
ending in the far-field and giving rise to the beam profiles shown
in Fig. 3 of the main paper, has been evaluated using a fully
quantum mechanical treatment. The diffraction effects are obviously
important in the propagation from near to far field, and in
particular in determining the presence or absence of destructive
interference effects at the vortex position that are essential in
our proposed scheme.

One may wonder if we can legitimately ignore the diffraction effects
in the other stages of our proposed setup. For the propagation
through the $q$-filter, we could assess directly the role of
diffraction in the case of homogeneous field, for which we have both
the exact solution of Pauli's equation (including diffraction) and
the classical trajectory solutions. We find that diffraction is only
relevant if we look at the electron beam very close to a focal
point, where its width becomes of the order of few nanometers. Away
from the focus, the ensemble of classical trajectories reproduces
perfectly well the quantum beam evolution. This result is not
surprising, as in the $q$-filter all length scales are 8--9 orders
of magnitude larger than the electron wavelength $\lambda=2\pi/k_c$,
which is of few picometers for energies of the order of 100 keV
(typical of electron microscopes). Therefore, we may safely assume
that diffraction can be ignored also in inhomogeneous $q$-filter
geometries, as long as no focusing occurs inside the filter. Our
classical ray-tracing simulations confirm that we avoid a focusing
inside the $q$-filter. It should be also noted that in designing and
analyzing electron-optics elements as those commonly used in
electron microscopy, it is quite standard to ignore quantum effects
and use classical ray tracing or geometrical optics.

The input and output apertures, being again about 7 orders of
magnitude larger than the electron wavelength, will also give
negligible diffraction effects (only some radial fringes extremely
close to the aperture edges are expected, which should not affect
the proper working of the proposed setup).

\subsection{Violation of Bohr's ``theorem''}
As reported in Ref.\ \cite{darrigol84}, few years after the
discovery of electron spin, several experiments were proposed (and
sometimes attempted) aimed at separating electrons according to
their spin, mainly by exploiting inhomogeneous magnetic fields
acting on the magnetic dipole associated with it (similar to the
Stern-Gerlach experiment). In 1929, Niels Bohr demonstrated a sort
of ``impossibility theorem'' \cite{darrigol84}, stating that no
Stern-Gerlach-like experiment (including Knauer's proposal, in which
an electric field was used to balance the Lorentz force) could be
used to separate electrons according to their spin by more than a
small fraction of the beam transverse extension. This was related to
the uncertainty principle and to the unavoidable orthogonal magnetic
field accompanying the required magnetic field gradient. A simple
explanation of this result is that the spin magnetic moment vanishes
in the classical limit $\hbar\rightarrow0$, so all attempts at
separating particles according to the magnetic-moment force
occurring in a magnetic field gradient give rise to a vanishing
separation in the classical limit, and hence a separation that is
comparable to quantum uncertainty. Actually, this statement has been
later proved not to be strictly valid (see, e.g., Refs.\
\cite{batelaan97,gallup01}), although it remains extremely hard to
overcome in practice.

Our approach is however fundamentally different from past attempts.
Indeed, we exploit the quantum nature itself of the electrons to
operate the spin separation (as such, our proposal bears some
similarities with a very recently proposed scheme for
spin-polarization in which electron diffraction is also exploited
\cite{mcgregor11}). As we have seen, the magnetic dipole force in
our device acts only in the azimuthal direction, which is not the
direction along which we separate the electrons. The torque
associated with this force imparts the OAM variation corresponding
to the azimuthal geometric phase. However, at the exit of our Wien
filter the electrons show no significant spatial separation
according to the spin. It is then the subsequent free propagation
and diffraction that acts to separate the electrons along the radial
direction, which is not the direction of the main field gradient. In
particular, an efficient radial separation is ensured by the
presence of a vortex at the beam axis for one spin component and not
for the other. In other words, we use the quantum wavy nature of the
electrons to separate them according to the spin, and hence our
method is not subject to Bohr's impossibility theorem. As a further
argument, even in the classical Stern-Gerlach approach Bohr's
theorem does not forbid the creation of spin-polarized components at
the edges of the beam, so that by suitable aperturing it should be
possible to generate spin-polarized beams, although with very low
efficiency. Our method is also based on selecting only an edge of
the beam, although this is done in the radial direction and the
efficiency is strongly improved by the vortex-related destructive
interference effects. Finally, we note that in our geometry the
unavoidable orthogonal magnetic field associated with the azimuthal
field gradient generating the dipole forces, which is the main
source of problems in the Stern-Gerlach approach to spin-polarized
electron separation, is already taken into account and corresponds
to the radial magnetic field of the quadrupole or hexapole
configuration. So its effect has been already fully considered in
our treatment.

%
\end{document}